# Identification of new assembly mode in the Heliconical Nematic Phase *via* Tender Resonant X-ray Scattering

Yu Cao[1,2], Jun Feng[2], Asritha Nallapaneni[2,3], Yuki Arakawa[4], Keqing Zhao[5], Georg H. Mehl[1,6], Feng Liu[1*], Chenhui Zhu[2*]

[1] State Key Laboratory for Mechanical Behavior of Materials, Shaanxi International Research Center for Soft Matter, School of Materials Science and Engineering, Xi'an Jiaotong University, Xi'an 710049, P. R. China

[2] Advanced Light Source, Lawrence Berkeley National Laboratory, Berkeley, CA 94720 USA

[3] Department of Polymer Engineering, University of Akron, Akron, OH 44325 USA

[4] Department of Applied Chemistry and Life Science, Graduate School of Engineering, Toyohashi University of Technology, Toyohashi, 441-8580 Japan

[5] College of Chemistry, Sichuan Normal University, Chengdu 610016, P. R. China

[6] Department of Chemistry, University of Hull, Hull HU6 7RX, U.K.

Correspondence should be addressed to C.Z. (email: chenhuizhu@lbl.gov) or to F.L. (email: feng.liu@xjtu.edu.cn)

**Abstract**

Helical structures are exciting and are utilized in numerous applications ranging from biotechnology to displays to medicine. Accurate description and understanding of resonance effects in helical structures provides crucial knowledge on molecular packing beyond positional ordering. We examined the manifestation of resonance effects in a nematic phase with heliconical structure, the so called twist bend nematic ($N_{TB}$) *via* tender resonant X-ray scattering (TReXS) at the sulfur K-edge. We demonstrate for the first time quantitatively that the energy dependence of the scattering peak in the $N_{TB}$ phase follows the energy dependence of the complex refractive indices measured by X-ray absorption. This allows us to identify a new self-assembly mode for specific sets of liquid crystal dimers in the $N_{TB}$ phase. We anticipate that new avenues in the exploration of complex orientational structures both in static as well as in dynamic modes induced by external stimuli will be pursued.

**Introduction**

Self-assembled hierarchical structures have emerged as an important class of advanced functional materials due to synergistic and often unique optical, mechanical, electrical and hydrophobic properties and found both in natural and artificial systems. Helical structures, if considered as a subset of hierarchical assemblies, are special, examples are DNA, forming the basis of life, cellulose in trees where helicity imbues mechanical strength, cholesteric liquid crystals (LCs) where the helical pitch defines the observed colour[1-3]. For technological applications of LCs precise measurement and full control of helicity and associated properties is crucial[4]. Hence the observation of a nematic like LC phase, found initially in chemically non-chiral dimeric molecules and characterized by a pitch ranging typically between 8-12 nm, has garnered significant attention in recent years[5-17], particularly as potential applications in electro-optical devices are discussed[18-21]. For these systems a short pitch helix was proposed, based on optoelectric studies[18,20]. Freeze fracture TEM (FFTEM)[22] and AFM[14,15] confirmed these proposals with directly observable data. Questions associated with surface induced structuring were addressed by NMR[13], Raman scattering[23] and conventional X-ray scattering[9,24] as well as carbon K-edge and Se K-edge X-ray resonant scattering[25-27]. Recently, chirality was directly measured using synchrotron circular dichroism of aligned samples[28]. Though basic features of this phase, assembly of molecules in nanoscale pitch helices, forming thus a pseudo-layer structure[29] with the mesogens at a tilt to the helix axis are now established, the nature of this phase is still discussed controversially. Existing materials do not behave fully in line with prevailing models[7,28,30,31] and alternatives are discussed and supported by experimental data and simulations[32]. Depending on the type of materials, organisation in duplexes, pentahedral or octahedral helices has been proposed; the precise measurements of the twisting and tilting of linear molecules in these heliconical structures is only at the beginning[33]. The somewhat contested nomenclature for the heliconical phase is not focus of this contribution, for simplicity we use the widely used term "$N_{TB}$".

In order to clarify the self-assembly behaviour and to move to a full understanding of the structure properties relationships, e.g. spatial variation of molecular orientation essential for the utilization of the existing properties of this phase and the rational design of advanced materials, precise measurement and description of the nanoscale helical organisation are required. Conventional X-ray scattering techniques that rely on spatial electron density fluctuations to provide structural details cannot provide essential information on complex bond orientation or molecular orientation. Resonant X-ray scattering, also known as anomalous X-ray scattering, overcomes the limitations associated with conventional X-ray scattering by taking advantage of the tunable, often enhanced, scattering power near elemental absorption edges. This technique has been mainly used to provide important structural information, such as counterion distribution in DNA/protein conjugates[34,35], morphology in multi-component copolymers[36], mean molecular orientations at the interfaces of polymer blends[37,38]. Recently, resonant soft X-ray scattering (RSoXS) has been demonstrated to be an unique and effective tool to directly probe periodic layer/molecular orientation variation in the bent-core B4 helical nanofilaments[39], $N_{TB}$ phase[25,27], blue phases[40], and other, so far not fully identified phases[41-43], based on the unique bond orientation sensitivity at the Carbon K-edge. Nevertheless, utilization of soft X-rays for structural examination is often associated with practical challenges: (1) due to the low penetration power of the soft X-rays a high vacuum environment for the samples is required and sample thickness are often limited to the submicron scale[44], and (2) the presence of multiple carbon atoms in a single organic molecule may pose a challenge to decipher local molecular-level interactions and packing. These complications can be circumvented by utilizing more penetrating tender X-rays (1 keV to 5 keV), which covers K-edges of elements such as Na, Mg, P, S, Cl, Si, K, Ti. Therefore, further development of tender resonant X-ray scattering (TReXS) offers great promise for the exploration of complex structures including biomaterials, battery materials, porous metal organic framework, with natural presence of these above elements. TReXS at the S K-edge was previously applied to discover smectic-C* liquid crystal variants[45,46], however, its applications to other materials have been very limited, one reason for which has been the difficulty in quantitative data interpretation and tensor-based scattering pattern prediction.

Here we address the aforementioned difficulty directly, through our investigations of a set of materials containing liquid crystal dimers with at least one thioether linkage using TReXS. Scattering peaks due to heliconical structure behave as purely resonant owing to pure bond orientation variation without electron density modulation. Meanwhile, peaks from layered structures (SmA) are partially resonant due to smectic layering, i.e. a nano-segregation of rigid molecular cores and flexible hydrocarbon tails exhibiting strong electron density modulations. Furthermore, we demonstrate that the dramatic intensity increase for the $N_{TB}$ phase scattering peak near S K-edge is driven by the strong energy dependence of complex molecular scattering factors, $f(E)=f_0+f'(E)+if''(E)$, measurable using X-ray absorption spectroscopy, with a precision significantly better than that computed from atomic form factor data base[44]. Finally, based on precise TReXS pitch temperature dependence, we propose a molecular packing model for asymmetric dimer. Our findings provide a novel route to reveal key structural information related to bond orientation in a broad class of natural/artificial hierarchical materials and provide a new idea about $N_{TB}$ phase to community.

**Results**

**Studied materials.** Two classes of liquid crystals were investigated and their phase sequences are listed in Fig. 1. The first one are the cyanobiphenyl (CB) based dimers with sulfur atoms present at the linking positions between the mesogenic groups and the central alkyl chains of molecules CBSC*n*SCB (*n* = 7) and CBSC*n*OCB (*n* = 5 and 7), exhibiting the $N_{TB}$ phase[47,48]. The second material (FBTBT) is newly synthesized according to a reported method[49] (Fig. 6) exhibiting a SmA phase and has sulfur atoms located in the aromatic cores. The well characterized molecule 8CB (4'-Octyl-4-biphenylcarbonitrile) is rod-shaped as FBTBT but contains no sulfur atoms and was used as a SmA reference.

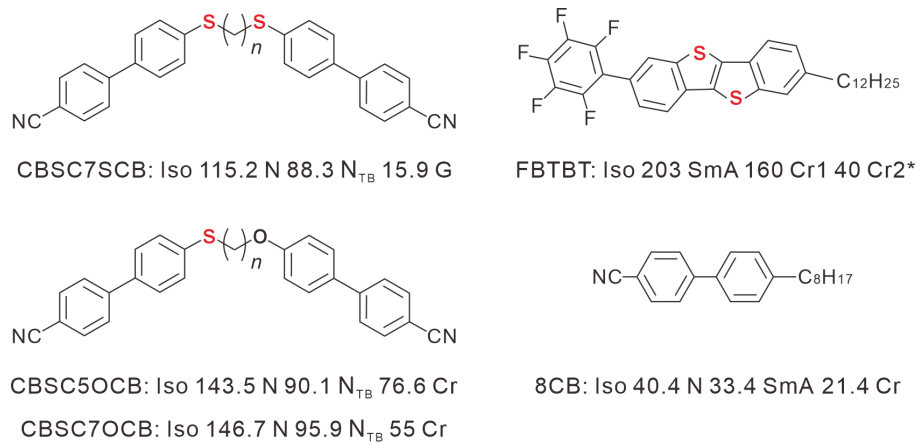

**Figure 1. Molecular structures and phase sequences.** The dimers CBSC*n*SCB and CBSC*n*OCB were reported to exhibit the $N_{TB}$ phase[47]. The newly synthesized sulfur-containing FBTBT exhibited SmA phase, and sulfur-free 8CB was used as a reference for the SmA phase, *n* is the number of repeating -$CH_2$- units.

**Resonant scatterings of liquid crystal and crystal phases.** The $N_{TB}$ phase of CBSC7SCB exhibited one single scattering peak at *q* = 0.72 nm$^{-1}$, which is only visible near the S K-edge (Fig. 2a) and corresponds to the helical pitch of the $N_{TB}$ phase. The helical pitch increased (decreased in the value of $q$) as the temperature increased towards the $N_{TB}$-N transition (Fig. 2b). A similar temperature dependence of the helical pitch was observed in other $N_{TB}$ compounds (Fig. 4). Remarkably, the peak intensity was observed to increase dramatically when the X-ray energy is increased towards the S K-edge (Fig. 2a and Supplementary Fig. 1). The peak intensity reduced abruptly right after the S K-edge, which was attributed to the increase in absorption of resonant atoms (Supplementary Fig. 5). All $N_{TB}$ materials examined in this study exhibited strong energy dependence of the scattering peak intensity (Supplementary Figs. 1 and 8). This observation is qualitatively similar to the one noticed in the $N_{TB}$ phase in CBC7CB[25], B4 helical nanofilaments in NOBOW[39] and three-dimensional blue phases[40]. However, the energy-dependence of such a 'forbidden' scattering peak originating from periodic bond orientation variation has not been accounted for quantitatively in any previous liquid crystal work and will be discussed below in comparison with the SmA case, where the scattering is expected to be dominated by smectic layering.

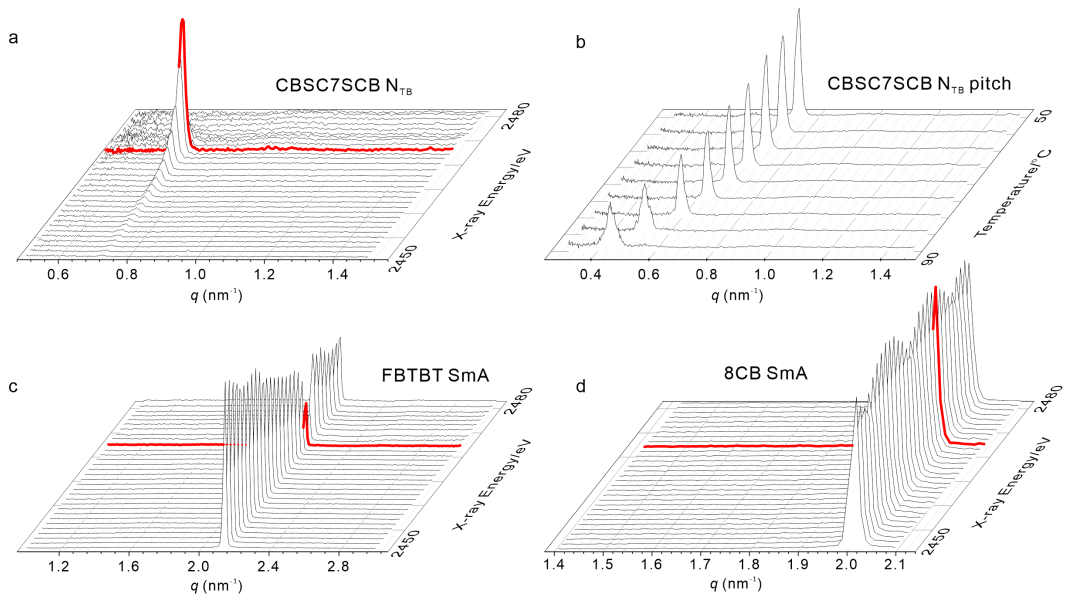

**Figure 2. Energy/temperature scan of $N_{TB}$ and SmA phases.** (a) TReXS Energy scan of the $N_{TB}$ phase of CBSC7SCB at 60 °C indicates the presence of an energy dependent resonant peak at $q = 0.72$ nm$^{-1}$; (b) Temperature scan of the $N_{TB}$ phase at $E = 2471$ eV upon cooling from 90 °C to 50 °C. Pitches decreased upon cooling from the nematic phase; (c), (d) TReXS Energy scans of the SmA phase of FBTBT (with sulfur atoms) at 195 °C and 8CB (without sulfur atoms) at room temperature, respectively. A sudden intensity dip is observed in the vicinity of S K-edge for FBTBT but not for 8CB. Thus considered partially resonant. Red thick lines correspond to the scattering at $E = 2741$ eV, slightly below S K edge.

The SmA phase of the sulfur-containing FBTBT exhibits one peak at $q = 2.14$ nm$^{-1}$ regardless of the X-ray energy (Fig. 2c), corresponding to a *d*-spacing ($= 2\pi/q$) of 2.93 nm. This value is very close to the extended molecular length (~3.10 nm), confirming that this peak has its origin from electron density modulation perpendicular to the plane of the layers. Interestingly, the peak intensity decreased noticeably at the S K-edge, which is clearly related to resonant sulfur atoms compared with reference 8CB (Fig. 2c, d) and fundamentally different from the dramatic increase in N$_{TB}$ peak at the S K-edge (Fig. 2a, c). For the crystal phase of CBSC7OCB and FBTBT several scattering peaks were detected at room temperature (Supplementary Fig. 2). Especially, there are two sharp peaks at $q = 2.04$ nm$^{-1}$ and $q = 4.08$ nm$^{-1}$ for CBSC7OCB crystal phase in a ratio of 1:2 representing smectic-like packing. Additionally, TReXS data for the crystal phase of CBSC7OCB contains one pure resonant peak at $q = 1.02$ nm$^{-1}$, which is exactly a double layer distance. Here we focus on the more representative energy-dependence behavior in the SmA and N$_{TB}$ phases.

**Theoretical explanation of resonant scattering.** We adopted the scattering intensity expression[34] represented in Eq. 1 to understand the nature of resonant scattering, wherein the complex scattering factor is given by *f(E)=f$_0$+f′(E)+if″(E)* and *v(q)* represents the spatial distribution of resonant species independent of X-ray energy. This expression reveals that the intensity measured near absorption edge consists of three parts: (1) the first term (*F$_0$$^2$(q)*) denotes the non-resonant intensity that is measured from atoms far from their absorption edge; (2) the second term (*2f′(E)F$_0$(q)v(q)*) corresponds to the cross term of the non-resonant and the resonant part (thus partially resonant), and scales linearly with *f′(E)*; Note that *f′(E)* is negative and accounts for the reduction in the effective number of electrons whereas the term *f$_0$+f′(E)* contributes to scattering (Supplementary Fig. 7); (3) the third term, *(f′$^2$(E) +f″$^2$(E))v$^2$(q)*, is based on pure resonant scattering and scales with the sum of *f′$^2$(E)* and *f″$^2$(E)*.

$$I_0(q) = F_0^2(q) + 2f'(E)F_0(q)v(q) + (f'^2(E) + f''^2(E))v^2(q) \qquad \text{Eq. (1)}$$

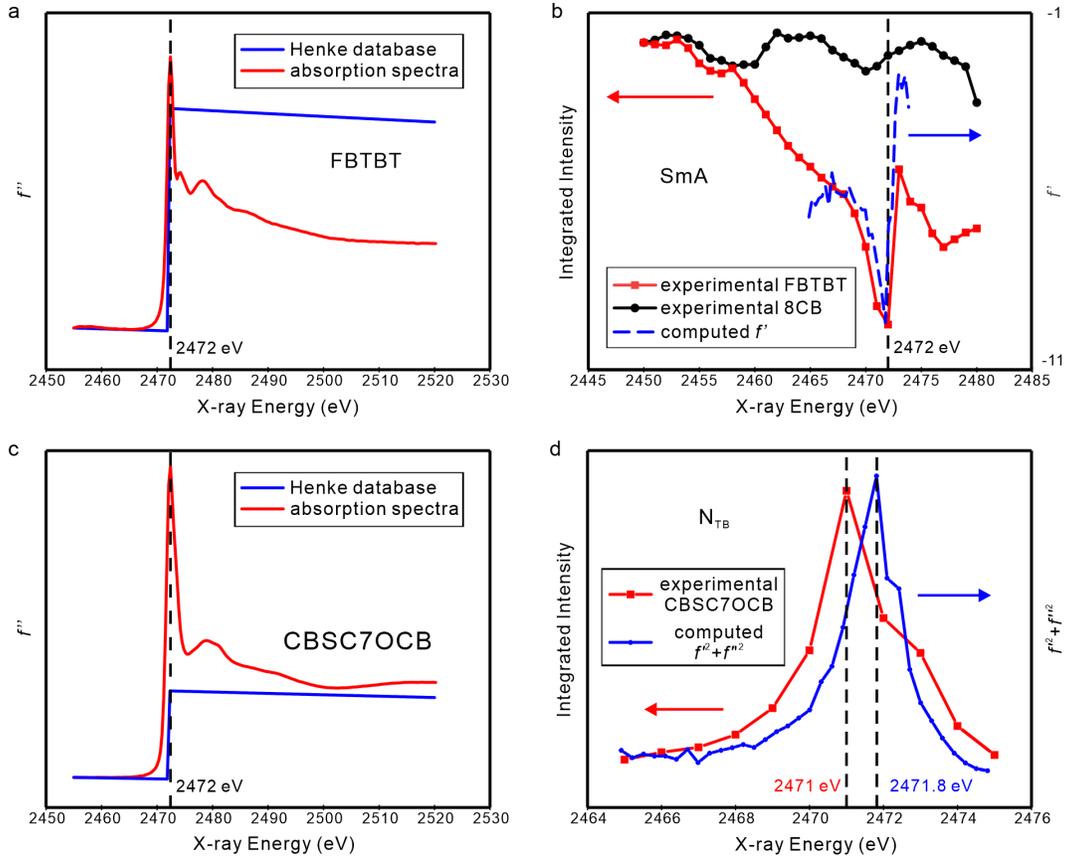

**Figure 3. Absorption spectra and comparison between computed and experimental scattering intensities.** (a) Imaginary part of the dispersion correction ($f''$) computed from experimental absorption spectra (red) taken in the SmA phase (at 195°C) and atomic form factor data base (blue) of FBTBT. Significant differences are observed around the sulfur absorption edge (2472 eV); (b) The integrated intensity of the SmA peak of 8CB (black), partially resonant SmA peak for FBTBT (red), and computed real part of the dispersion correction ($f'$) (blue dash line) *vs* X-ray energy. For comparison, all of intensities of 8CB are scaled by a multiplier constant which was obtained from the intensity of FBTBT and 8CB at $E = 2450$ eV. A dramatic dip around the S K-edge can be found for FBTBT. The computed $f'$ shows a similar dip at the S K-edge suggesting that the intensity dip in FBTBT arises from resonant sulfur atoms; (c) Computed imaginary part of dispersion correction ($f''$) from absorption spectra (red) and atomic form factor data base (blue) of CBSC7OCB at 90°C; (d) Experimental integrated intensity of $N_{TB}$ peak of CBSC7OCB (red) and computed scattering contrast ($f'^2+f''^2$) (blue) *vs* X-ray energy. Experimental data reaches its maximum at 2471 eV with strong energy dependence. Intensity increased dramatically, i.e. by a factor of > 23 from 2465 to 2471 eV. Both experimental and computed results exhibit similar trends around absorption edge except for a slight shift attributed to the instrumental resolution limitation.

First, we checked the SmA case with the above expression. Based on the experimental data provided in Fig. 2, integrated peak intensity for the SmA and $N_{TB}$ phases as a function of X-ray energy were plotted as shown in Fig. 3. As 8CB contains no sulfur, the observed small fluctuation in integrated intensity is attributed to background fluctuations due to beamline optics (Fig. 3b). The SmA phase in FBTBT shows an additional intensity dip near the S K-edge, which is clearly due to the sulfur atoms present in FBTBT. Given a large non-resonant $F_0^2(q)$ term in the SmA phase of FBTBT (Fig. 3b), the cross term in Eq. (1) contains the leading resonant contribution and therefore should be the main perturbation to the measured scattering intensity. To calculate the contribution of the partial resonant term quantitatively, X-ray absorption spectra near the S K-edge were measured (Supplementary Fig. 5) to obtain the imaginary dispersion correction, $f''(E)$ from FBTBT (Fig. 3a) following $\mu = 2\rho N_a r_e \lambda f''/m_a$, where $\mu$ is the attenuation length, $\rho$ is the density, $N_a$ is the Avogadro constant, $r_e$ is the classical electron radius, $m_a$ is the atomic molar mass, and $\lambda$ is the X-ray wavelength. Evidently, the experimental imaginary dispersion correction $f''(E)$ near the S K-edge (Figs. 3a,c and Supplementary Fig. 6) differs significantly from the simulated ones based on the atomic form factor database[44]. In organic molecules, both molecular orbital hybridization and specific local chemical environments affect the details in near-edge absorption spectra. The corresponding dispersive component, $f'(E)$, in the SmA phase of FBTBT, was calculated from $f''(E)$ using the Kramers-Kronig relation[50], and that overlaps with the measured scattering intensity, $I(E)$, reasonably well (Fig. 3b). This is a strong indication that the peak intensity in the SmA case follows the adopted expression as long as the experimental refractive indices (dispersion and absorption), rather than the simulated, are used for the near-edge region.

Next, we discuss the energy dependence of peak intensity for the $N_{TB}$ case, which has never been accounted quantitively for the $N_{TB}$ peak or the superlattice peak of Smectic-C* variants[45]. In the $N_{TB}$ phase, non-resonant SAXS data showed no peak at $q = 2\pi/p$, where $p$ corresponds to the full helical pitch, suggesting that the non-resonant term $F_0^2(q)$ in Eq. 1 is negligible and the electron density modulation from the helical structure is minimal. Fig. 3d also indicates that the observed scattering peak near the S K-edge is due to pure resonance, which agrees with the $N_{TB}$ structural model that exhibits periodic bond orientation variation from a screw axis. As reported previously, the precise description of bond orientation sensitivity and structure factor calculation typically requires a treatment of a second rank tensorial form factor[40,51-53], which can be modelled as a 3-by-3 matrix. This is in the simplest case a diagonal matrix with only two unequal parameters, i.e. the scattering factors parallel, $f_{para}$, and perpendicular, $f_{perp}$, to the molecular long axis or specific chemical bonds of interest. Here, we assume a fixed anisotropy in the scattering factor, i.e. a constant ratio of $f_{perp}/f_{para}$ independent of X-ray energy. We focus on the general quantitative energy-dependence of such an anisotropy, which is responsible for the pure resonant peaks from periodic orientation variation. Complex dispersion correction, $f'(E)$ and $f''(E)$ were computed from the measured absorption spectra (Supplementary Figs. 5-7, Supplementary Table 1). Fig. 3d indicates that the computed $f'^2(E)+f''^2(E)$ curve matches reasonably well with the results of measured peak intensity as a function of energy (except for a slight shift in the peak position), thus validating the hypothesis that the energy-dependence of the $N_{TB}$ peak intensity basically follows the energy dependence of complex molecular scattering form factors (Supplementary Fig. 8). We note that a similar expression has been adopted for calculating compositional contrast in multicomponent polymeric systems[54] but has not been extended to systems with pure orientational ordering such as the helical $N_{TB}$ liquid crystals[25,55], B4 helical nanofilaments[39], blue phases[40], B2 phases[56] or Smectic-C alpha phases[45,57,58].

**Discussion**

This in-depth understanding of the TReXS data offers us a chance to decipher the local molecular packing in the $N_{TB}$ phase better. As discussed above, it is surprising that the pitch for both asymmetric dimers (CBSC7OCB, CBSC5OCB) $p_{as}$ was found to be significantly larger than that of the symmetric dimer (CBSC7SCB) $p_s$, see Fig. 4. Clearly noticeable is the decrease of the helical pitch with lowering the temperature from the nematic/$N_{TB}$ transition temperature. For CBSC7SCB this follows the trend reported for a number of cyanobiphenyl based dimers investigated by RSoXS[25,40] and is in line too with results for difluorosubstituted dimers using Se K-edge resonant scattering[26]. This has been associated with a decreasing number of twisted conformations of the molecules at higher temperatures. In other words, at high temperatures, due to temperature induced fluctuations the helical ordering unwinds, the pitch is larger and decreases with lowering the temperature. For previously investigated systems a pitch ranging typically between 8-12 nm has been reported[25,26,40,55], typically the lengths of 2.5-3 extended dimers, equivalent to 5-6 monomeric units, which is in line with CBSC7SCB. The results for the materials CBSC5OCB and CBSC7OCB follow this general trend, however, they behave qualitatively different to materials reported earlier. The pitch is significantly larger than for other dimeric materials, reaching values of 18.4 nm for CBSC7OCB and 14.8 nm for CBSC5OCB respectively. As the difference in the overall molecular length between CBSC7OCB (3.32 nm), CBSC7SCB (3.16 nm) is very small, differences in molecular size cannot account for this qualitatively different behaviour. Calculated differences in the opening angles of the bent shaped molecules of 119° for CBSC7OCB, 94° for CBSC7SCB indicate an argument for different behaviour. However, we note that for CBC7CB an opening angle of 111° is calculated[48], that for CBC9CB is 102°[59] and for systems involving a nonyl spacer pitches between 8-12 nm were also detected[55]. All this indicates that the pitch is not a simple function of bond angles between spacers and aromatic cores. This has already been discussed for the dimer/trimeric systems used for the structurally related system CBC6OCB[27].

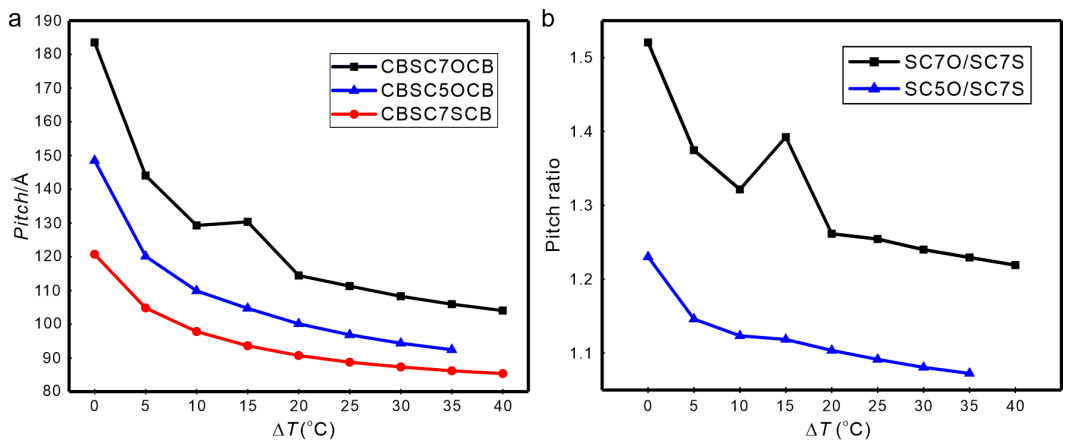

**Figure 4. Temperature dependence of helical pitches and their ratios.** (a) Pitch variation vs reduced temperature ($\Delta T$ = $T_{NTB-N}$-$T$) in the $N_{TB}$ phases. $\Delta T$ denotes temperature difference below the transition temperature from the nematic to $N_{TB}$ phase upon cooling. The pitch vs $\Delta T$ follows similar trend for all materials; (b) Pitch ratio between asymmetric dimers (CBSC7OCB and CBSC5OCB) and the symmetric dimer (CBSC7SCB) vs reduced temperature.

The small angle peaks for the asymmetric dimers such as CBSC7OCB (~1.25 nm) associated with molecular clustering are slightly smaller than the monomer lengths ~1.68 nm measured in the all-trans form using Materials Studio starting from central carbon in the central spacer to terminal cyano group, suggesting a partially intercalated structure[60,61]. Temperature dependent change was found to be minimal, as observed for other systems[47]. As illustrated in Fig. 4b, the ratio ($p_{as}/p_s$) of the pitches ($p_{as}$) of the asymmetric dimers and pitch ($p_s$) of symmetric CBSC7SCB are plotted as a function of the reduced temperature ($\Delta T=T_{NTB-N}-T$). Upon cooling, the pitch ratio $p_{as}/p_s$ gradually decreased and reached a limit far from phase transition temperature, indicating extra pitch decrement for asymmetric dimers. Here, a geometric model was constructed accounting a different mode of packing for asymmetric CBSC5OCB and CBSC7OCB in Fig. 5. With similar molecular shape and intercalation, our model suggests asymmetric dimers are packed in form of symmetric 'supramolecular tetramers' to generate larger helical pitch. The rotation between dimers in 'supramolecular tetramers' accounts for the extra pitch decrement.

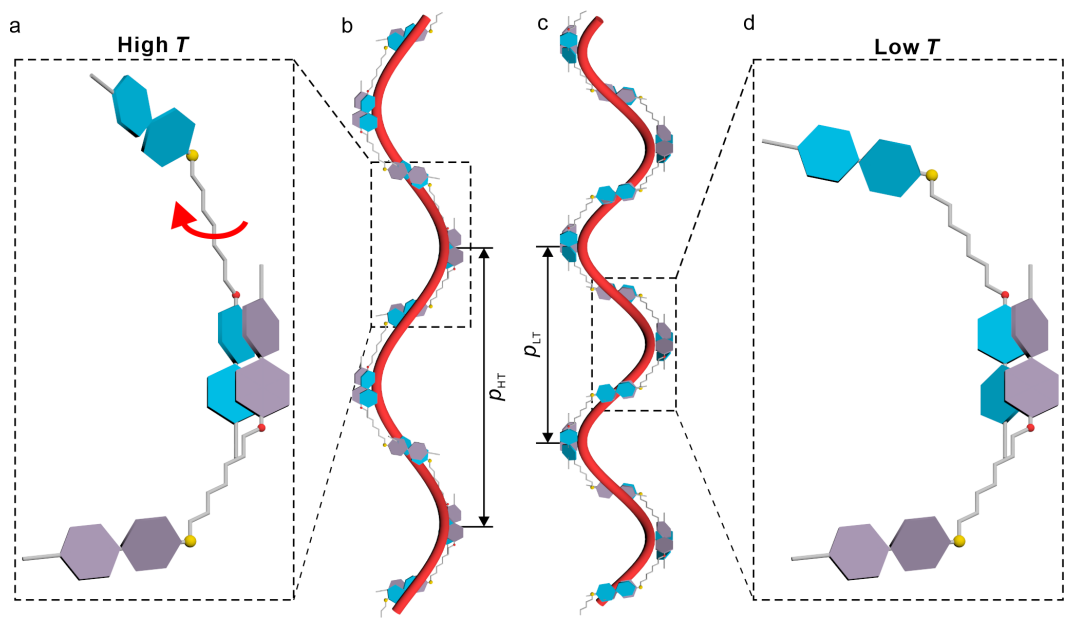

**Figure 5. Schematic molecular arrangement of the $N_{TB}$ phase by 'supramolecular tetramers'.** (a), (d) Distinct 'supramolecular tetramers' models at high and low temperatures, respectively. The red arrow indicates the rotation direction of blue dimer along intercalated biphenyl axis; (b), (c) The $N_{TB}$ phase packing model with intercalated biphenyl groups where red helices indicate the helix sense of 'supramolecular tetramers'. The twist in 'supramolecular tetramers' introduces extra pitch decrement as shown in Supplementary Fig. 9.

Plotting the reduced correlation length, in other words, the number of helical stacks as a function of the reduced temperature shows that the investigated behaviour is quite similar for all investigated systems, 10 °C below the transition the correlation length is equivalent to about 24-28 helical stacks for all materials (see Supplementary Fig. 10b). This is consistent with our model of 'supramolecular tetramers' formed by asymmetric dimers that are packed helically, similar to symmetric dimers. Subsequently we considered the crystal phase of asymmetric dimer CBSC7OCB. With help of two peaks in the small angle region and the pure resonant peak interpreted as bilayer distance, we reconstructed the electron density map with molecular packing (Supplementary Fig. 11). Determined from TReXS, the single layer distance is around 3.08 nm, which suggests an intercalated structure as in Supplementary Fig. 11. The resulted anticlinic molecular packing resembles that formed by 9-(3BEP)2[62]. There the crystal phase also indicates the similar packing as our proposed $N_{TB}$ model; based on the results discussed above, the structure would be an $N_{TB(int)}$ phase.

In summary, we have explained that TReXS is an enormously powerful and arguably unique technique to measure short pitch helical systems. We demonstrated this by identifying the first example of an $N_{TB}$ phase made up of associated dimers forming a 'supramolecular tetramer'; alternatively, this phase could be addressed as $N_{TB(int)}$ or $N_{PT(int)}$[30] using an alternative nomenclature. We anticipate that this methodology can be readily applied to obtain new insights of the self-assembly of the mysterious polarization modulated $SmA_{PFmod}$ phase in bent-core liquid crystals[63], the twist grain boundary smectics[64] and other emerging complex hierarchical structures. With quick experimental absorption spectroscopy measurement, model-dependent resonant X-ray scattering pattern can be computed in advance, which will greatly speed up the process of materials discovery.

**Methods**

**Synthesis.** The compound was synthesized in-house according to the synthetic procedure shown in Fig. 6. Details of the synthesis and the analytical data of intermediates and final compound are given in the accompanying Supplementary Information.

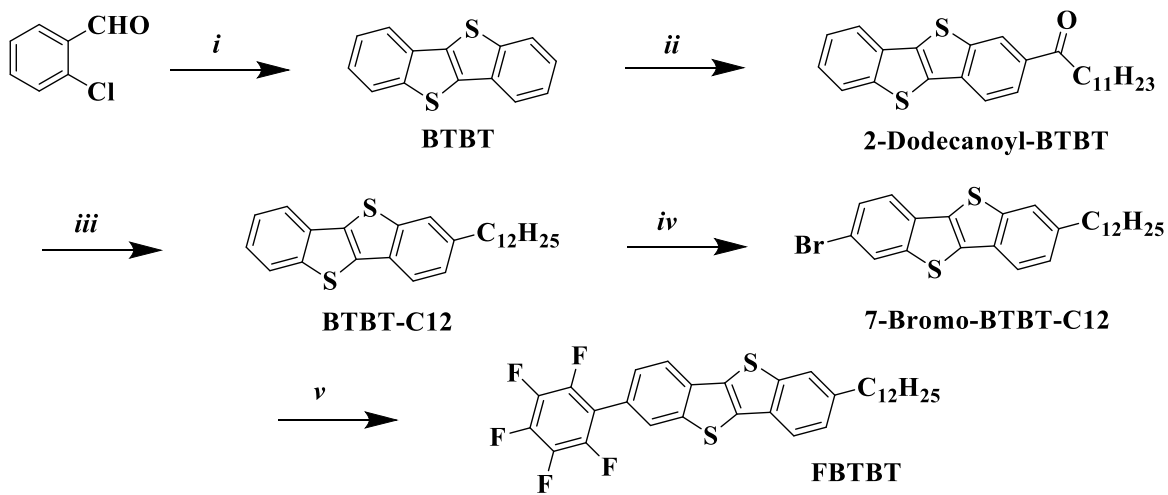

**Figure 6. Synthesis of compound FBTBT.** Reagents and conditions: (*i*) NaSH, NMP, 180 °C, 12h; (*ii*) $C_{11}H_{23}$COCl, AlCl$_3$, -36 °C, 2h; (*iii*) KOH, NH$_2$-NH$_2$, 220 °C, 28h; (*iv*) Br$_2$, CH$_2$Cl$_2$, rt, 2h; (*v*) C$_6$F$_5$B(OH)$_2$, Pd(PPh$_3$)$_4$, P(*t*-Bu)$_3$, CsF, Ag$_2$O, DMF, 100 °C, 12h.

**Tender resonant X-ray scattering.** TReXS measurement was performed at the beamline 5.3.1 at the Advanced Light Source, Lawrence Berkeley National Laboratory. The X-ray beam energy was tuned around the S K-edge, 2472 eV, with a channel cut double-bounce silicon (111) monochromator. A two-dimensional Pilatus detector (300K, Dectris, Inc.) was used to collect the scattering patterns, which were subsequently converted to one-dimensional line profiles using the Nika software package[65]. The scattering patterns were viewed with the Xi-Cam interface[66] at the beamline. The sample-detector distance was tuned between 488 mm and 250 mm to access relevant *q* range. The beam centers and the sample-to-detector distances were calibrated using both silver behenate and 8CB smectic A.

**X-ray absorption.** X-ray absorption spectroscopy measurement was performed at the beamline 5.3.1 at the Advanced Light Source, Lawrence Berkeley National Laboratory. The X-ray beam energy was tuned around the S K-edge, 2472 eV, with a channel cut double-bounce silicon (111) monochromator. X-ray absorption spectra were measured with a photodiode in transmission mode. The practical energy resolution is about 1 eV. To reduce air attenuation, the sample chamber was kept in a helium gas environment.

**Acknowledgements**

We thank Dr. Eric Gullikson at CXRO, LBNL and Dr. Howard Padmore at ALS, LBNL, for helpful discussions. We acknowledge use of Beamlines 5.3.1 of the Advanced Light Source supported by the Director of the Office of Science, Office of Basic Energy Sciences, of the U.S. Department of Energy under contract no. DE-AC02-05CH11231. The work was also supported by National Natural Science Foundation of China (No. 21774099, 21761132033 and 51603166), Science and Technology Agency of Shaanxi Province (2016KW-050 and 2018KWZ-03) and the 111 Project 2.0 (BP2018008). Y.C. also thanks China Scholarship Council (CSC) for providing financial support (201706280170).

**Author contributions:**

F.L. and C.Z. initialized the project. Y.C. carried out the TReXS measurement under the guidance of F.L. and C.Z. and with help of J.F.. Y.A. and K.Z. offer the materials. Y.C. analyzed the experimental data. G.M. and A.N. helped in analyzing and discussing the results. Y.C., F.L. and C.Z. wrote the paper with inputs from all authors.

**Additional information**

**Supporting Information**. Additional information on TReXS, absorption spectra, atomic form factors, calculations of molecular form factors from measured absorption spectra, integrated peak intensity *vs* beam energy, are available free of charge via the Internet.

**Competing interests.** The authors declare no competing interests.

# Supplemental Information

# Identification of new assembly mode in the Heliconical Nematic Phase *via* Tender Resonant X-ray Scattering

Yu Cao[1,2], Jun Feng[2], Asritha Nallapaneni[2,3], Yuki Arakawa[4], Keqing Zhao[5], Georg H. Mehl[1,6], Feng Liu[1*], Chenhui Zhu[2*]

[1] State Key Laboratory for Mechanical Behavior of Materials, Shaanxi International Research Center for Soft Matter, School of Materials Science and Engineering, Xi'an Jiaotong University, Xi'an 710049, P. R. China

[2] Advanced Light Source, Lawrence Berkeley National Laboratory, Berkeley, CA 94720 USA

[3] Department of Polymer Engineering, University of Akron, Akron, OH 44325 USA

[4] Department of Applied Chemistry and Life Science, Graduate School of Engineering, Toyohashi University of Technology, Toyohashi, 441-8580 Japan

[5] College of Chemistry, Sichuan Normal University, Chengdu 610016, P. R. China

[6] Department of Chemistry, University of Hull, Hull HU6 7RX, U.K.

# Synthesis of FBTBT

**(1) Benzothieno[3,2-b]benzothiophene (BTBT)**

To a 100 mL flask, 2-chlorobenzaldehyde (5 g, 35.6 mmol), N-methyl-2-pyrrolidone (NMP, 15 mL), sodium hydrosulfide hydrate (NaSH, 8 g, 0.11 mol) were added and heated at 80°C for 1 h; then heated at 180°C for 12 h with stirring. After the reaction was finished, which was monitored by thin-layer chromatography (TLC), the reaction mixture was poured into saturated aqueous $NH_4Cl$ solution. The deposit was collected by filtration and washed by ethanol. Recrystallization from toluene resulted in a yellow crystals BTBT (1.92g, 45%).

$^1$H NMR ($CDCl_3$, TMS, 400 MHz) δ: 7.86–7.92 (m, 4H, 4ArH), 7.37–7.47 (m, 4H, 4ArH).

**(2) 2-Dodecanoyl-BTBT**

In a 250 mL flask, BTBT (2 g, 8.32 mmol) and dichloromethane (200 mL) were added and protected under argon, and was cooled to −17°C. Then $AlCl_3$ (3 g, 22.5 mmol) was added and cooled to −36°C. Dodecanoyl chloride (2 g, 9.14 mmol) was added slowly and stirred at −36°C for 2 h, then room temperature for 1 h. After reaction was finished, cold water and methanol were added. The extraction by dichloromethane was further purified by column chromatography (Silica, petroleum/dichloromethane/toluene, 3:1:1, v). It was recrystallized from toluene to give white crystals 2-dodecanoyl-BTBT (3 g, 85%).

$^1$H NMR ($CDCl_3$, TMS, 400 MHz) δ: 8.31(d, J = 6.8 Hz, 1H, ArH), 7.92 (s, 1H, ArH), 7.52 (d, J = 7.2 Hz, 1H, ArH), 7.44–7.47 (m, 2H, ArH), 2.58 (d, J = 7.2 Hz, 2H, $CH_2$), 1.76 (t, J = 6.8 Hz, 2H, $CH_2$), 1.28 (s, 16H, $CH_2$), 0.89 (t, J = 6.8 Hz, 3H, $CH_3$).

**(3) BTBT-C12**

To a 100 mL flask, 2-dodecanoyl-BTBT (200 mg, 0.47 mmol), KOH (139.4 mg, 2.48 mmol), 2-ethoxyethanol (20 mL), and hydrazine hydrate (80%, 0.7 mL) were added and heated at 120°C for 1 h, and heated at 220°C for 28 h. The reaction was monitored by TLC. Cooling the mixture resulted in solid crystals, and recrystallization from toluene yielded white solid 2-dodecyl-BTBT (189.5 mg, 98%).

$^1$H NMR ($CDCl_3$, TMS, 400 MHz) δ: 7.90 (d, J = 8.0 Hz, 1H, ArH), 7.87 (d, J = 8.0 Hz, 1H, ArH), 7.79 (d, J = 8.4 Hz, 1H, ArH), 7.72 (s, 1H, ArH), 7.45 (t, J = 7.2 Hz, 1H, ArH), 7.38 (t, J = 7.2 Hz,

1H, ArH), 7.29 (d, J = 8.0 Hz, 1H, ArH), 2.76 (t, J = 8.0 Hz, 2H, CH$_2$), 1.61–1.78 (m, 2H, CH$_2$), 1.26–1.33 (m, 18H, CH$_2$), 0.88 (t, J = 7.2 Hz, 3H, CH$_3$).

**(4) 7-Bromo-BTBT-C12**

To a 50 mL flask was added 2-dodecyl-BTBT (800 mg, 1.96 mmol) and dichloromethane (16 mL). The solution of Br$_2$ (313.6 mg) in CH$_2$Cl$_2$ (5mL) was added and stirred at room temperature for 3 h. The reaction was monitored by thin-layer chromatography. The solid was filtered out and recrystallized from toluene and yielded a white solid, 7-bromo-2-dodecyl-BTBT (181.3 mg, 76%).
$^1$H NMR (CDCl$_3$, TMS, 400 MHz) δ: 8.04 (s, 1H, ArH), 7.78 (d, J = 8.0 Hz, 1H, ArH), 7.71 (t, J = 6.0 Hz, 2H, ArH), 7.53–7.56 (m, 1H, ArH), 7.28 (d, J = 8.4 Hz, 1H, ArH), 2.76 (t, J = 7.6 Hz, 2H, CH$_2$), 1.65–1.73 (m, 2H, CH$_2$), 1.25–1.34 (m, 18H, CH$_2$), 0.88 (t, J = 6.4 Hz, 3H, CH$_3$).

**(5) FBTBT**

To an argon protected 25 mL reaction tube, 7-bromo-2-dodecyl-BTBT (147 mg, 0.3 mmol), C$_6$F$_5$B(OH)$_2$ (70mg, 0.33mmol), CsF (91.2 mg, 0.6 mmol), Ag$_2$O (45 mg, 0.36 mmol), and Pd(PPh$_3$)$_4$ (41.2 mg, 0.036 mmol) were added. Then the solution of P(*t*-Bu)$_3$ (9.1 mg, 0.045 mmol) in DMF (3 mL) was injected, and stirred at 100 °C for 12 h. Silica gel TLC was used to monitor the reaction (eluted by petroleum ether: CH$_2$Cl$_2$ = 2: 1, v). It was cooled, the solid deposit collected, washed by water (5 mL x 2) and CH$_2$Cl$_2$ (5 mL x 2), purified by silica gel column chromatography with elution of toluene and recrystallization in toluene-petroleum ether to get the target molecule (88.1mg, 51%).
$^1$H NMR (CDCl$_3$, TMS, 400MHz) δ: 7.97 (d, J = 8.0 Hz, 2H, ArH), 7.83 (d, J = 8.0 Hz, 1H, ArH), 7.74 (s, 1H, ArH), 7.43-7.50 (m, 1, ArH), 7.24-7.32 (m, 1H, ArH), 2.71 (t, J = 7.6 Hz, 2H, CH$_2$), 1.69-1.83 (m, 2H, CH$_2$), 1.26-1.35 (m, 18H, CH$_2$), 0.88 (t, J = 7.2 Hz, 3H, CH$_3$).
HRMS (m/z) Calcd. for C$_{32}$H$_{31}$F$_5$S$_2$: 574.18; Found: 574.1785.

# Experimental data

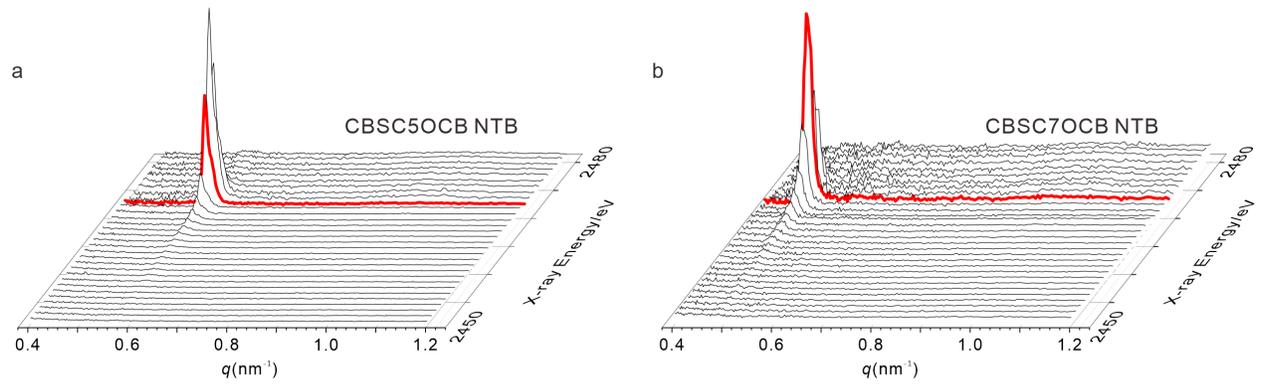

**Supplementary Figure 1. Energy scan of the $N_{TB}$ phase in asymmetric dimers.** (a) CBSC5OCB and (b) CBSC7OCB. All $N_{TB}$ peaks show strong energy dependence near S K-edge. Red thick lines correspond to the scattering at $E = 2741$ eV, slightly below S K edge.

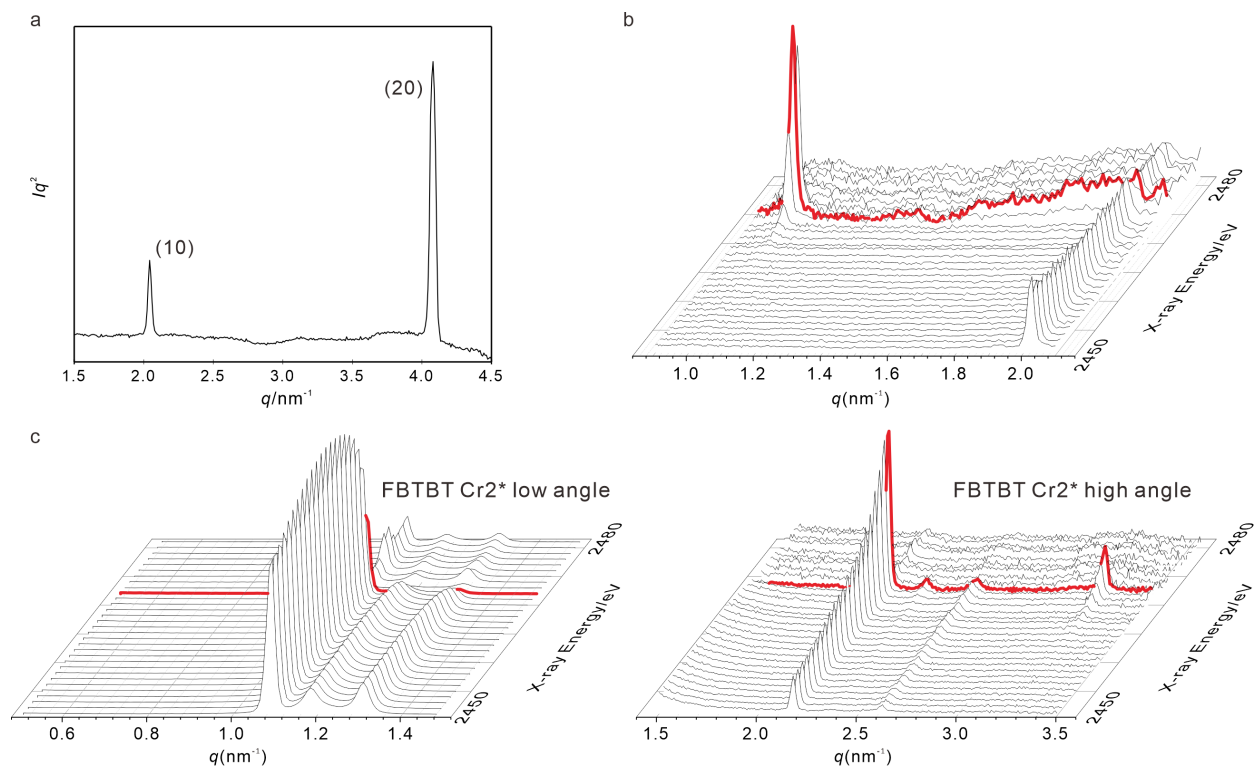

**Supplementary Figure 2. SAXS and TReXS energy scan of two crystal phases from asymmetric dimer and rod-like molecule.** (a) SAXS diffractogram of CBSC7OCB at crystal phase. To avoid resonant effect, the beam energy was tuned to 3keV. The two scattering peaks are in a ratio of 1:2 indicating a smectic-like phase; (b) TReXS energy scan of crystal phase of CBSC7OCB at room temperature. The resonant signal represents double-layer distance which exhibits a feature of anticlinic packing; (c) TReXS energy scan of crystal phase of FBTBT at room temperature. Low and high angle regions were shown separately so that distinct features can be observed clearly. Peaks at low angle region were visible even far away from the S K-edge with significant intensity dip as partially resonant SmA peak. The peak at $q$ = 3.25 nm$^{-1}$ is visible only in the vicinity of S K-edge and considered as pure resonant peak. The peak at $q$ = 2.17 nm$^{-1}$ exhibited combined features of both partially resonant and pure resonant peaks. Red thick lines correspond to the scattering at $E$ = 2741eV, slightly below S K edge.

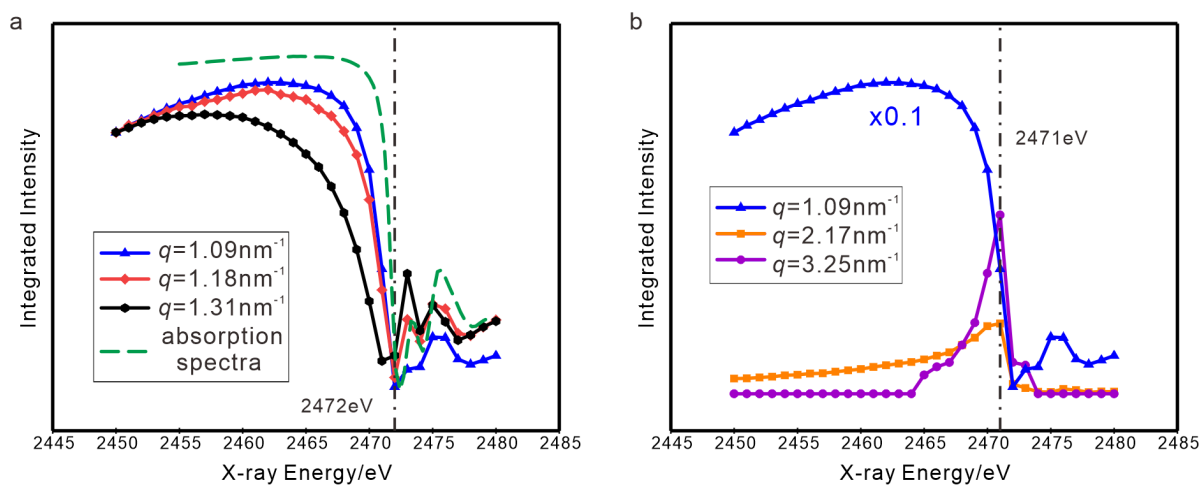

**Supplementary Figure 3. Energy dependence of different scattering peaks of FBTBT crystalline phase.** (a) The integrated intensity of three non-resonant peaks and the measured absorption spectra *vs* X-ray energy. The non-resonant peaks and absorption spectra exhibit similar trends; (b) Plot comparing the integrated intensities of partially resonant peak (blue), pure resonant peak (purple) and resonant enhanced peak (orange). Note that for the pure resonant peak (purple), the intensity is set to be zero before 2464eV and after 2474eV because it was nearly visible experimentally.

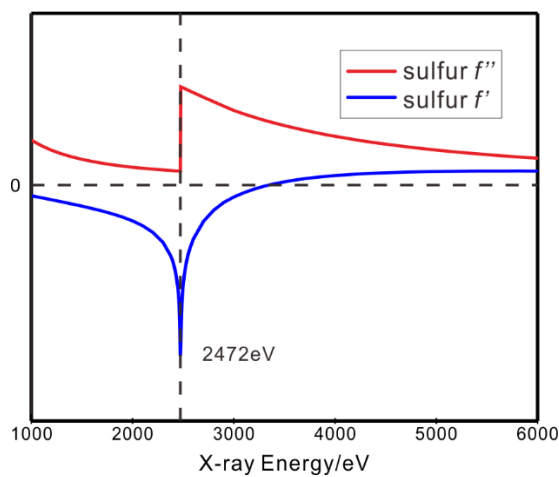

**Supplementary Figure 4.** Real ($f'$) and imaginary ($f''$) part of dispersion correction of sulfur atom from theoretical approximation developed by Cromer and Liberman[1,2].

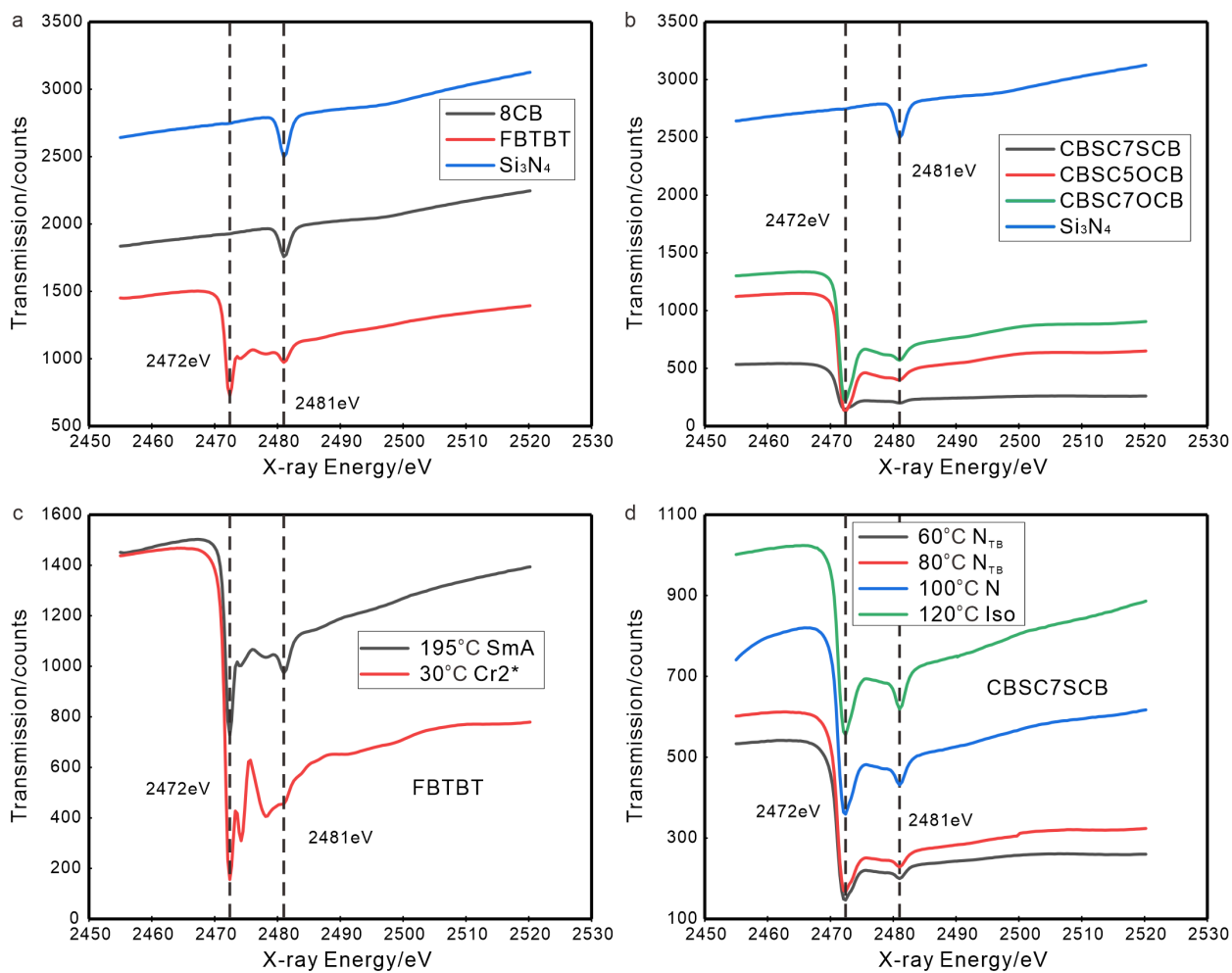

**Supplementary Figure 5. Absorption spectra of all materials studied in different phases.** (a) SmA phase of FBTBT, 8CB and amorphous $Si_3N_4$ (reference). Absorption at S K edge (2472 eV) is prominent and the presence of another dip at 2481 eV is likely to be arising from beamline optics, i.e. the crystal glitches as reported by Gerrit Van Der Laan and Bernard T. Thole[3]; (b) $N_{TB}$ phase of CBSC7SCB, CBSC7OCB, CBSC5OCB and amorphous $Si_3N_4$ (reference); (c) Absorption spectra of SmA and crystalline phases of FBTBT. The spectra are different for different phases indicating that absorption spectral can reflect structural information; (d) Absorption spectra of CBSC7SCB at different temperatures exhibiting $N_{TB}$, N and isotropic phases. Significant differences on absorption effect can be observed.

Liquid crystal samples were sandwiched between two pieces of Si$_3$N$_4$ membranes (Norcada) and the absorption spectroscopy data were collected in transmission mode with a pin diode. To compute absorption coefficient, the film thickness is needed to acquire attenuation length $\mu$ from Beer-Lambert Law $I(x) = I_0 e^{-\mu x}$, where $I_0$ is the incoming X-ray intensity before the sample, $\mu$ is attenuation length and $x$ is sample thickness. In our case, we assumed that the attenuation length at 2455eV (before S K-edge) is close to that from the Center for X-ray Optics, which is based on Henke atomic scattering factors database[2]. The calculated sample thicknesses are listed in Table S1. The absorption coefficient (Supplementary Fig. 7) and dispersion correction (Supplementary Fig. 8) as a function of X-ray energy were then computed and plotted.

**Supplementary Table 1.** Liquid crystal sample thickness calculated from absorption spectroscopy

| Sample name | CBSC7SCB | CBSC7OCB | CBSC5OCB | FBTBT | 8CB |
|---|---|---|---|---|---|
| Film thickness /$\mu$m | 95.7 | 41.4 | 49.6 | 27.4 | 23.5 |

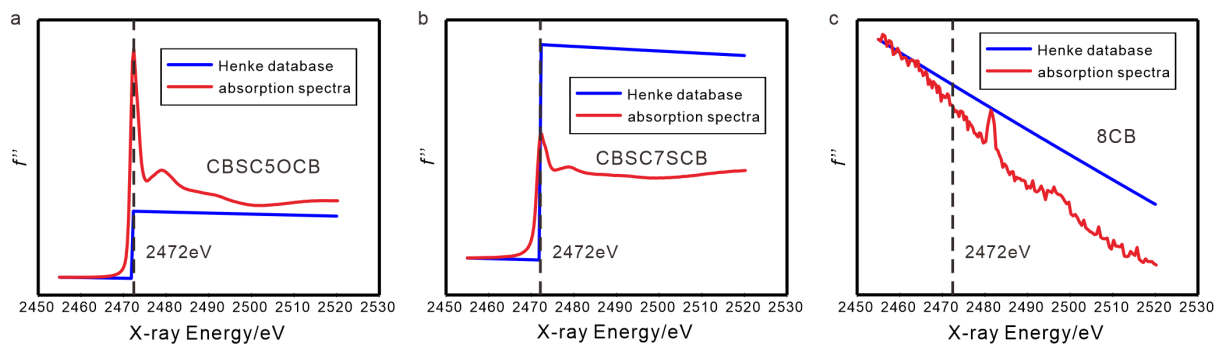

**Supplementary Figure 6.** Imaginary part of dispersion correction $f''$ vs X-ray energy computed from experimental absorption spectra and Henke atomic scattering factors database. The minor fluctuation around 2481 eV is visible even in empty samples, and is attributed to the background from beamline optics.

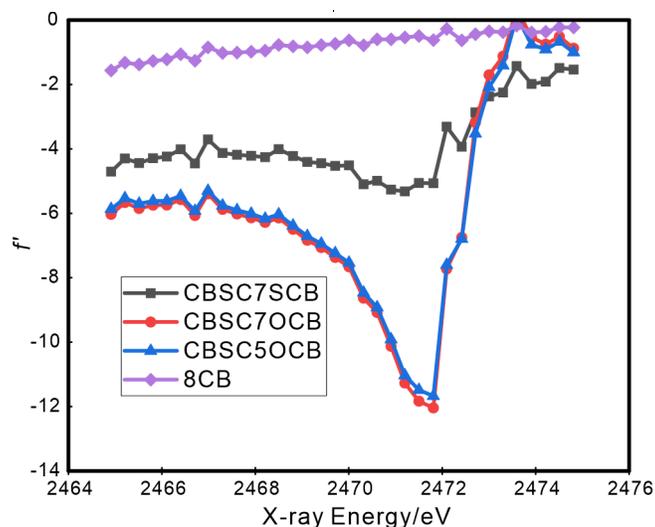

**Supplementary Figure 7.** Real part of dispersion correction *f'* *vs* X-ray energy calculated from *f''* by Kramers-Kronig relation[4].

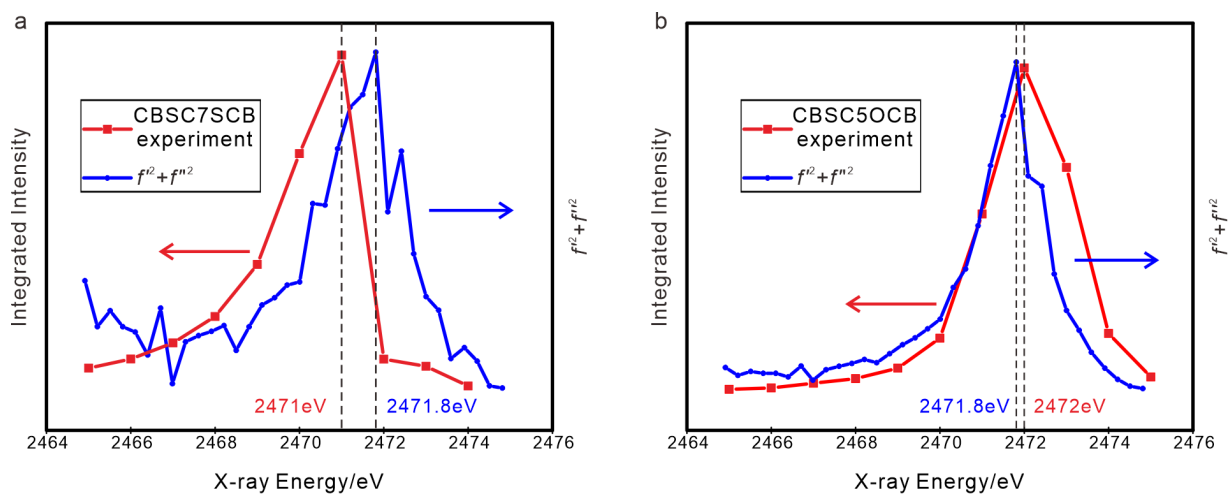

**Supplementary Figure 8. Integrated intensity of $N_{TB}$ peaks and computed scattering contrast ($f'^2 + f''^2$) *vs* X-ray energy of different dimers.** (a) CBSC7SCB and (b) CBSC5OCB. Both the plots suggest that the experimental peak intensity as a function of X-ray energy followed the same trend as the computed scattering contrast, which is essentially determined by the complex molecular scattering factor. This emphasizes the fact that $N_{TB}$ peak is purely resonant.

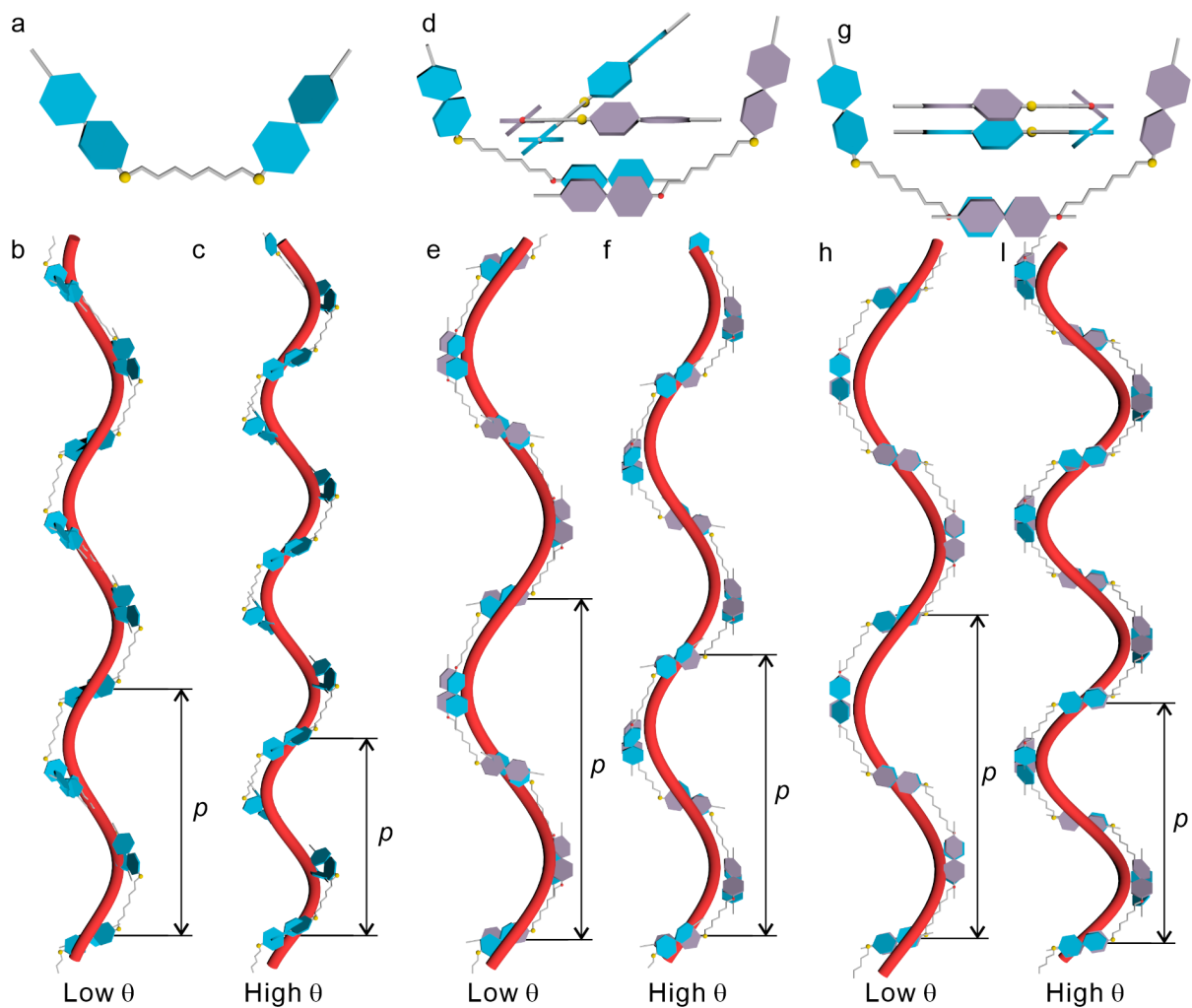

**Supplementary Figure 9. Schemitic representation of dimer shape and packing in N$_{TB}$ phase of dimers.** (a)-(c) symmetric dimer, (d)-(f) high temperature 'supramolecular tetramer' and (g)-(i) low temperature 'supramolecular tetramer'. N$_{TB}$ model at low $\theta$ applied tilt angle $\theta$ around 10º while high $\theta$ around 40º according to birefringence results[5]. By comparing (f) and (i), the extra twist along biphenyl axis decreases helical pitch, which induces the extra decrement of helical pitch in asymmetric system.

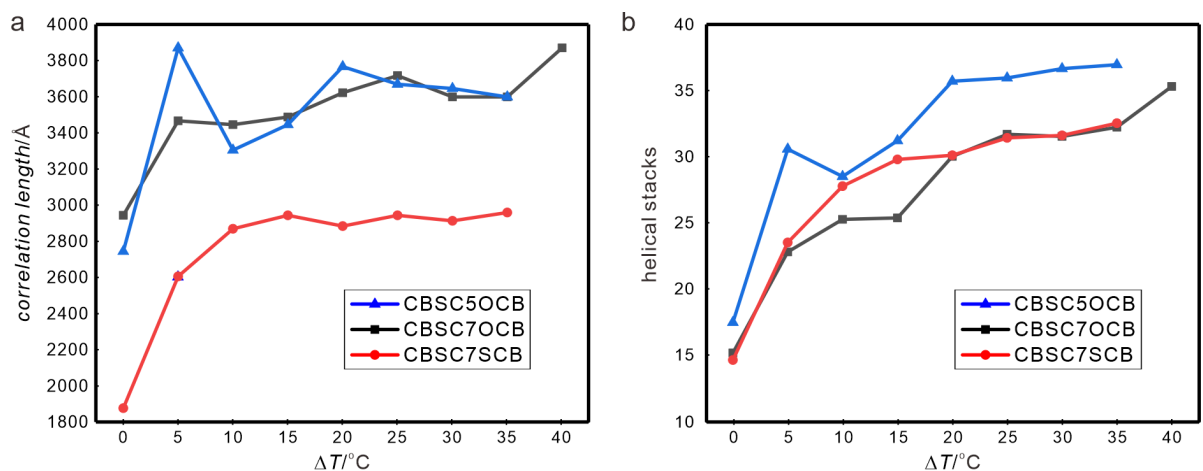

**Supplementary Figure 10. Computed correlation length and helical stacks from TReXS data.** (a) Correlation length and (b) helical stacks calculated from TReXS temperature scan at 2471 eV. Correlation length $\lambda$ is roughly calculated by $\lambda = 2\pi/\text{FWHM}$ while helical stacks is obtained by $\lambda/p$.

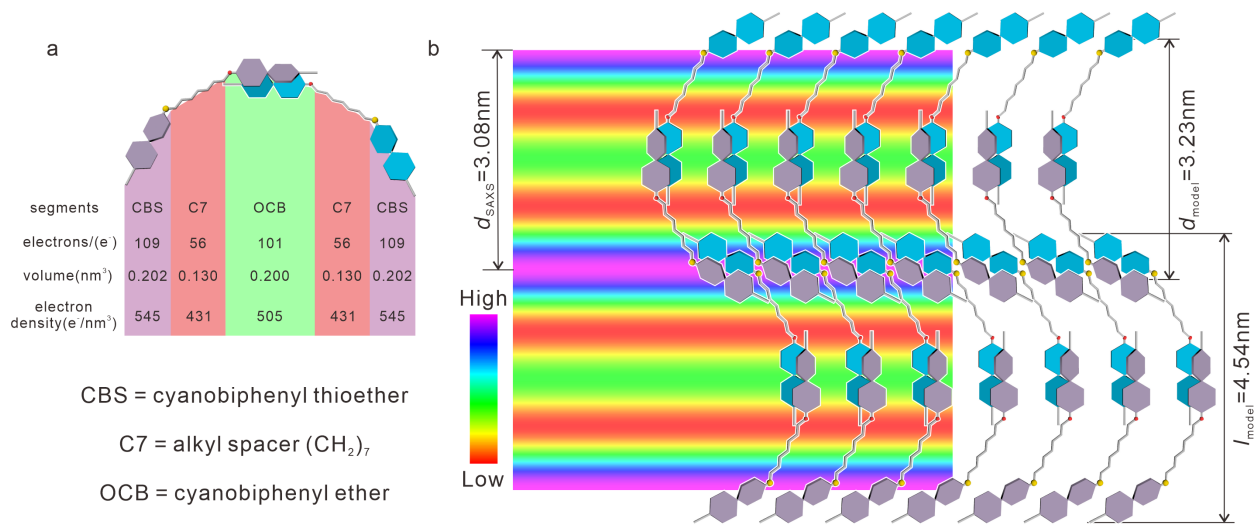

**Supplementary Figure 11. Deciphering crystal structure of asymmetric dimer by combining SAXS and TReXS data.** (a) Electron density distribution of CBSC7OCB, a dimer is shown to match the electron density map; (b) Reconstructed electron density map with superimposed molecular model of CBSC7OCB. The ED map is reconstructed by SAXS scattering intensities in Fig. S2(a)[6]. The molecular size is in line with calculated ED layer distance.

## Supplementary References